\documentclass{elsart}
\usepackage{natbib}
\usepackage{graphicx}
\begin{document}
\runauthor{Y.~Aritomo}
\begin{frontmatter}
\title{Fusion hindrance and roles of shell effects in superheavy
mass region}
\author{Y.~Aritomo}

\address {Flerov Laboratory of Nuclear
Reactions, JINR, Dubna, Russia}

\begin{abstract}

We present the first attempt of systematically investigating the
effects of shell correction energy for a dynamical process, which
includes fusion, fusion-fission and quasi-fission processes. In the
superheavy mass region, for the fusion process, shell correction
energy plays a very important role and enhances the fusion
probability when the colliding partner has a strong shell structure.
By analyzing the trajectory in three-dimensional coordinate space
with the Langevin equation, we reveal the mechanism of the
enhancement of the fusion probability caused by `cold fusion
valleys'. The temperature dependence of shell correction energy is
considered.

\end{abstract}
\end{frontmatter}




\section{Introduction}

In a heavy-ion fusion reaction, with increasing atomic numbers of
the target and projectile, it becomes more difficult to produce a
compound nucleus owing to the strong Coulomb repulsion and strong
dissipation forces. In the superheavy mass region, this difficulty
is more pronounced. Although the mechanism of the fusion-fission
reaction in the heavy-mass region is not clear, generally we
recognize the existence of fusion hindrance
\cite{hana74,pete76,brun79}.

The dynamical aspects of the fusion-fission mechanism have been
investigated as a time evolution of the nuclear shape with friction.
In the 1980s, a mean trajectory calculation in the nuclear
deformation space using a stochastic equation, known as the
extra-push model, was performed by Swiatecki and coworkers
\cite{swia81,bloc86}. In the superheavy-mass region, however, even
if substantial extra push energy is supplied, it is very difficult
for the mean trajectory to reach the spherical region owing to a
strong dissipation. It is necessary to apply fluctuation-dissipation
dynamics \cite{ari04}.

The simplified calculation that included the one-dimensional
diffusion model was applied to cold fusion reactions \cite{swit05}.
With one adjustable parameter, the results showed a good agreement
with many experimental data, not only maximum cross sections, but
also the optimum energies and widths of the excitations functions.
In the present study, we concentrate to discuss the fusion process
using the more realistic model.

Fusion hindrance is mainly caused by the macroscopic properties of
the colliding partner \cite{ari05I}. On the contrary, fusion is
enhanced by the deformation and shell structure of the nuclei, which
we call fusion enhancement \cite{mits00,nish00,sato02,ikez03}. In
experiments, especially in the case of a cold fusion reaction, these
advantages are used to synthesize superheavy nuclei
\cite{hofm00,mori03}. To understand the fusion mechanism clearly, it
is better to treat separately fusion hindrance and fusion
enhancement, that is to say, the macroscopic and microscopic
aspects, respectively.


In our previous study \cite{ari05I}, we focused, in particular, on
fusion hindrance, in the superheavy mass region. We have presented
the origin of the fusion hindrance systematically by conducting a
trajectory calculation in three-dimensional coordinate space. To see
the fundamental mechanism of the fusion hindrance, we employed the
potential energy of the liquid drop model, and analyzed the mean
trajectory. From the behavior of the trajectory, we can understand
the mechanism that the fusion probability decreases exponentially as
the $Z$ number of the fused system increases. It is concluded that
the fusion hindrance in a system with $Z$ greater than 102 is
explained by considering the fragment deformation to be an important
factor.


Considering aspects of the fusion enhancement, we focus on the
influence of the shell effect. The nuclear structure of the
projectile-target combinations is related to the touching
probability, but it also influences the dynamics from the touching
point to the compound nucleus in the superheavy mass region
\cite{sato02,ikez03}. In our approach \cite{ari04,ari05,ari05I}, as
parameters related to the microscopic effects, in principle we can
introduce the shell correction energy on the potential energy
surface and the transport coefficients calculated using the
microscopic model \cite{ivan97,hofm97,yama97}. In the former case,
we can see 'cold fusion valleys' in the potential energy surface,
which have been suggested in references \cite{gra99,gra00}. It is
said that these valleys lead to the enhancement of the fusion
probability. Here, we investigate precisely how the trajectory is
affected by shell correction energy, and how the fusion probability
is changed.


In fact, shell correction energy depends on the nuclear temperature.
In a dynamical process, the nuclear temperature changes owing to the
dissipation of the kinetic energy of the relative motion. Therefore,
we must discuss the fusion process taking into account the time
evolution of the potential energy surface, which is controlled by
nuclear friction, since the dissipation depends on the nuclear
friction. We clarify the effect of the temperature-dependent shell
correction energy in the fusion process.


For clarity, in this paper, we consider the 'dynamical process' as
the whole process from the contact point to the spherical nucleus or
re-separated one. It includes the fusion, quasi-fission and deep
quasi-fission processes, which are defined in reference
\cite{ari04}.


In section~2, we briefly explain our framework for the study and the
model. We precisely investigate the effects of cold fusion valleys
in section~3. In this section, using the potential energy surface of
the full shell correction energy, we discuss the mechanism of the
enhancement of the fusion probability by cold fusion valleys. In
section~4, taking into account the temperature dependence of shell
correction energy, we discuss the fusion process. In section 5, we
present a summary and further discussion to clarify the reaction
mechanism in the superheavy mass region.


\section{Model}

Using the same procedure as described in reference \cite{ari04} to
investigate the dynamical process, we use the
fluctuation-dissipation model and employ the Langevin equation. We
adopt the three-dimensional nuclear deformation space given by
two-center parameterization \cite{maru72,sato78}. The three
collective parameters involved in the Langevin equation are as
follows: $z_{0}$ (distance between two potential centers), $\delta$
(deformation of fragments) and $\alpha$ (mass asymmetry of the
colliding nuclei); $\alpha=(A_{1}-A_{2})/(A_{1}+A_{2})$, where
$A_{1}$ and $A_{2}$ denote the mass numbers of the target and the
projectile, respectively.

The parameter $\delta$ is defined as $\delta=3(a-b)/(2a+b)$, where
$a$ and $b$ denote the half length of the axes of ellipse in the $z$
and $\rho$ directions, respectively as expressed in Fig.~1 in
reference \cite{maru72}. We assume that each fragment has the same
deformations as the first approximation. $\delta$ is related to the
deformation parameter $\beta_{2}$, which is familiar with us, as
\begin{equation}
\beta_{2}=\frac{\delta}{\sqrt{\frac{5}{16\pi}}(3-\delta)}.
\end{equation}

The neck parameter $\epsilon$ is the ratio of the smoothed potential
height to the original one where two harmonic oscillator potentials
cross each other. It is defined in the same manner as reference
\cite{maru72}. With $\epsilon < 1$, the surface of two fragments
shows the smooth curve at the connecting point of them. On the other
hand, in the case of $\epsilon=1$, the two fragments are connected
with a sharp point like a top of cone. In the present calculation,
$\epsilon$ is fixed to be 1.0, so as to retain the contact-like
configuration more realistically for two-nucleus collision.

The multidimensional Langevin equation is given as
\begin{eqnarray}
\frac{dq_{i}}{dt}&=&\left(m^{-1}\right)_{ij}p_{j},\nonumber\\
\frac{dp_{i}}{dt}&=&-\frac{\partial V}{dq_{i}}
                 -\frac{1}{2}\frac{\partial}{\partial q_{i}}
                   \left(m^{-1}\right)_{jk}p_{j}p_{k}
                  -\gamma_{ij}\left(m^{-1}\right)_{jk}p_{k}\nonumber
                  +g_{ij}R_{j}(t),\\
\end{eqnarray}
where a summation over repeated indices is assumed. $q_{i}$
denotes the deformation coordinate specified by $z_{0}$, $\delta$
and $\alpha$. $p_{i}$ is the conjugate momentum of $q_{i}$. $V$ is
the potential energy, and $m_{ij}$ and $\gamma_{ij}$ are the
shape-dependent collective inertia parameter and dissipation
tensor, respectively. A hydrodynamical inertia tensor is adopted
in the Werner-Wheeler approximation for the velocity field, and
the wall-and-window one-body dissipation is adopted for the
dissipation tensor \cite{bloc78,nix84,feld87}. The normalized
random force $R_{i}(t)$ is assumed to be  white noise, {\it i.e.},
$\langle R_{i}(t) \rangle$=0 and $\langle R_{i}(t_{1})R_{j}(t_{2})
\rangle = 2 \delta_{ij}\delta(t_{1}-t_{2})$. The strength of
random force $g_{ij}$ is given by $\gamma_{ij}T=\sum_{k}
g_{ij}g_{jk}$, where $T$ is the temperature of the compound
nucleus calculated from the intrinsic energy of the composite
system.
The potential energy is defined as

\begin{equation}
V(q,l,T)=V_{LD}(q)+\frac{\hbar^{2}l(l+1)}{2I(q)}+V_{SH}(q,T),
 \label{vt1}
\end{equation}
\begin{equation}
V_{LD}(q)=E_{S}(q)+E_{C}(q),
\end{equation}

\begin{equation}
V_{SH}(q,T)=E_{shell}^{0}(q)\Phi (T),
\end{equation}
where $I(q)$ is the moment of inertia of a rigid body at deformation
$q$. $V_{LD}$ and $V_{SH}$ are the potential energy of the
finite-range liquid drop model and the shell correction energy
taking into account the temperature dependence. $E_{shell}^{0}$ is
the shell correction energy at $T=0$. The temperature dependent
factor $\Phi (T)$ is discussed in the section 4.1. $E_{S}$ and
$E_{C}$ denote a generalized surface energy \cite{krap79} and
Coulomb energy, respectively. The centrifugal energy arising from
the angular momentum $l$ of the rigid body is also considered. The
detail is explained in reference \cite{ari04}. The intrinsic energy
of the composite system $E_{int}$ is calculated for each trajectory
as

\begin{equation}
E_{int}=E^{*}-\frac{1}{2}\left(m^{-1}\right)_{ij}p_{i}p_{j}-V(q,l,T),
\end{equation}

where $E^{*}$ denotes the excitation energy of the compound nucleus.
$E^{*}$ is given by $E^{*}=E_{cm}-Q$, where $Q$ and $E_{cm}$ denote
the $Q-$value of the reaction and the incident energy in the
center-of-mass frame, respectively. At $t=0$, each trajectory starts
from the contact point which is defined as $R_{touch}=R_{p}+R_{t}$,
where $R_{p}$ and  $R_{t}$ are the radii of projectile and target,
respectively.

As discussed in reference \cite{ari04}, we assumed that the kinetic
energy of the relative motion does not dissipate during the
approaching process. However, to investigate the effect of the
energy dissipation during the approaching process, we introduce a
parameter $E_{diss}$ which is the intrinsic energy converted from
the relative kinetic energy before both nuclei touch. The discussion
on this issue is given in section 4.3. At the contact point, the
initial velocity is directed only in the $-z$ direction, and the
components of the initial velocities along the $\delta$ and $\alpha$
directions are both assumed to be zero \cite{ari04,ari052}.





\section{Effects of shell correction energy in fusion process}

In our previous study \cite{ari05I}, we focused on the fusion
hindrance from the macroscopic point of view. We employed the
potential energy of the liquid drop model and analyzed the mean
trajectory excluding the final term on the right-hand side of
Eq.~(2) and used $\Phi(T)=0$ in Eq.~(5). We presented the origin of
the fusion hindrance systematically.

Here, we investigate the fusion enhancement from the microscopic
point of view. The nuclear structure of the projectile-target
combinations is related to the touching probability, but it also
influences the dynamics from the contact point to the compound
nucleus \cite{sato02,ikez03}. We focus on the dynamics taking into
account the shell effect after both nuclei touch. Within our model
\cite{ari04,ari05,ari05I}, we introduce the potential energy surface
with the shell correction energy.

\begin{figure}
\centerline{
\includegraphics[height=.60\textheight]{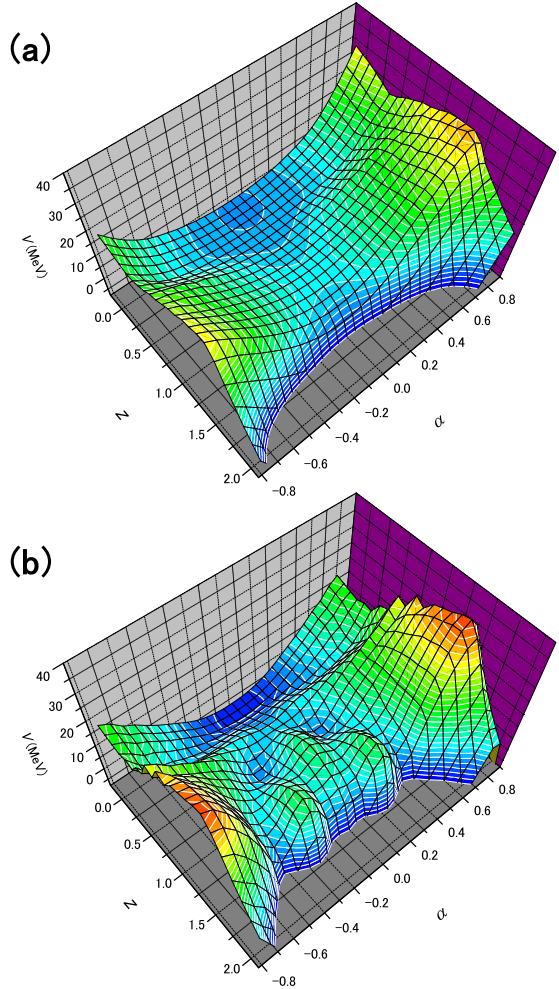}}
  \caption{Potential energy of the liquid drop model $(a)$ and with shell correction
   energy $(b)$ for
$^{292}114$ in the $z-\alpha$ space $(\delta=0)$. The calculation is
done using the two-center shell model code \cite{suek74,iwam76}.}
\label{1-3dim}
\end{figure}

Figure \ref{1-3dim} shows the potential energy surface of the liquid
drop model $(a)$ and with shell correction energy $(b)$ for
$^{292}114$ in the $z-\alpha$ space $(\delta=0)$, which is
calculated using the two-center shell model code
\cite{suek74,iwam76}. When we consider the shell correction energy
on the potential energy surface, we can see pronounced valleys, that
lead to the compound nucleus. The valleys are called 'cold fusion
valleys' \cite{gra99,gra00}. It is said that these valleys enhance
the fusion probability.

We can see such shell structure in another calculation
\cite{moll97}. Single particle level diagrams in reaction
$^{70}$Zn+$^{208}$Pb were calculated by a macroscopic-microscopic
model. It shows that the entrance-channel fragment-shell effects
remain far inside the touching point.

In this section, we discuss the effect of shell correction energy on
the dynamical process using trajectory calculation.

\begin{figure}
\centerline{
\includegraphics[height=.30\textheight]{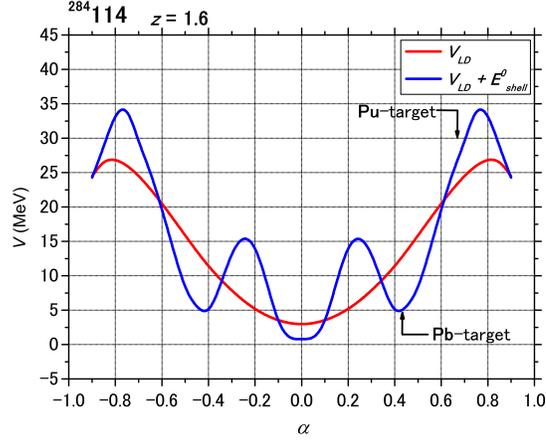}}
  \caption{Potential energy of $V_{LD}$ and $V_{LD}+E^{0}_{shell}$ for
$^{284}114$ at $z=1.6$, which are denoted by the red and blue lines,
respectively. The calculation is done using the two-center shell
model code. The combination of the Pb-target and Pu-target are
indicated.}
\end{figure}

\subsection{Effect of cold fusion valleys on fusion process}

As our first attempt, we employ the potential energy of the liquid
drop with the full shell correction energy, which corresponds to the
potential energy surface at $T=0$ MeV. This potential energy is
represented by $V_{LD}+V_{SH}(q,T=0)$, or $V_{LD}+E^{0}_{shell}$ by
Eq.~(5). We mainly discuss head-on collision, that is to say, the
process for $l=0$, because the general mechanism of the trajectory's
behavior does not change essentially in any of the angular momentum
cases, although the potential landscape changes with angular
momentum.


\begin{figure}
\centerline{
\includegraphics[height=.60\textheight]{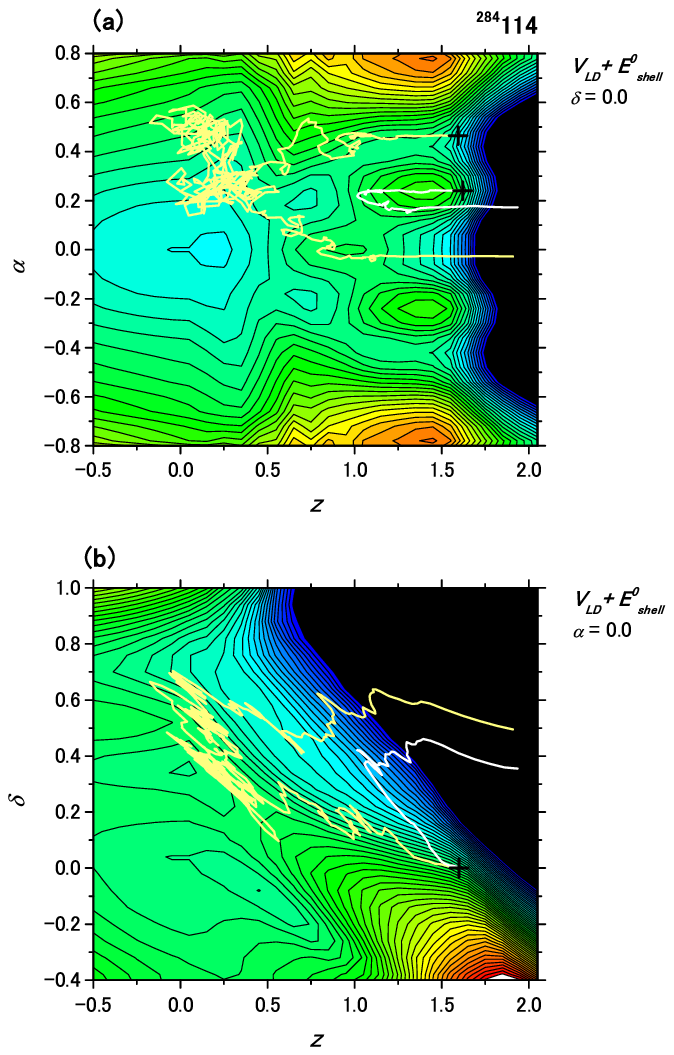}}
  \caption{Sample trajectories projected onto the $z-\alpha$ plane
at $\delta=0.0$ (a) and the $z-\delta$ plane at $\alpha=0.0$ (b) of
$V_{LD}+E^{0}_{shell}$ for $^{284}$114. The light yellow and white
lines denote the trajectories which start at $\alpha=0.46$ and 0.24
at $E^{*}=20$ MeV. Symbols are given in the text.}
\end{figure}


\begin{figure}
\centerline{
\includegraphics[height=.60\textheight]{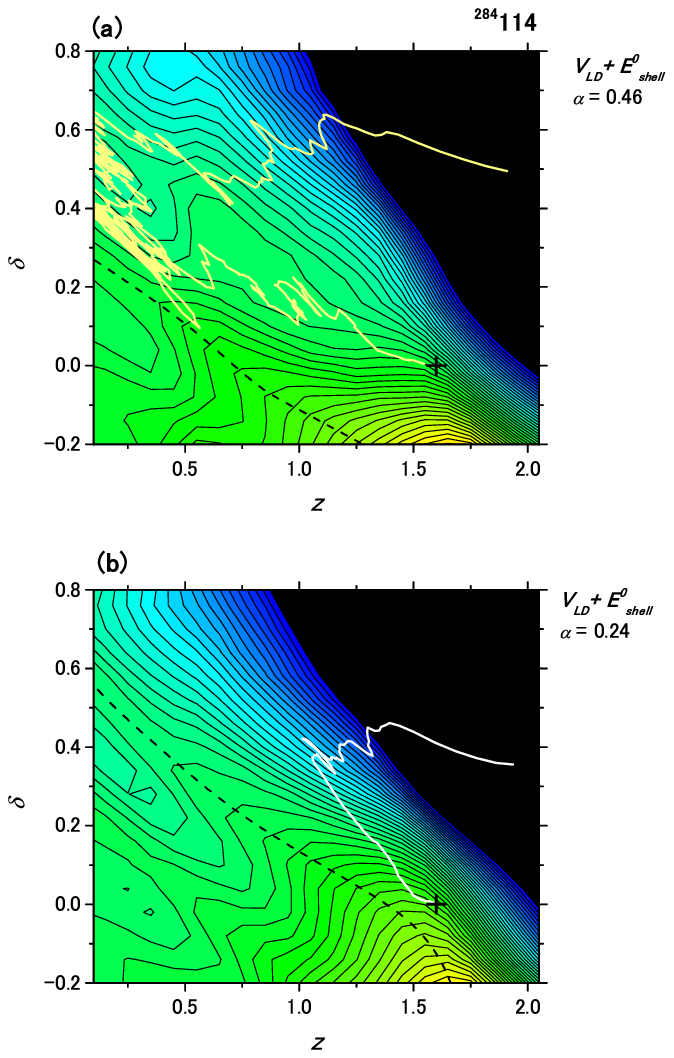}}
  \caption{The trajectories in Fig.~3(b) are projected on to the $z-\delta$ plane
  near the contact point. (a) $\alpha=0.46$ and (b) $\alpha=0.24$. The
dashed line denotes the ridge line of the potential energy surface.}
\end{figure}

Figure 2 shows the potential energies $V_{LD}$ and
$V_{LD}+E^{0}_{shell}$ for $^{284}114$ at $z=1.6$, which the
potential energies correspond to a location near the contact point.
The red and blue lines denote $V_{LD}$ and $V_{LD}+E^{0}_{shell}$,
respectively. The shell correction energy in the reaction system
where the target corresponds to Pb is significant, and is indicated
in Fig.~2. Actually, a Pb target is chosen in a cold fusion
reaction, in order to suppress the excitation energy of a compound
nucleus \cite{oga75}.



To investigate the effects of the cold fusion valleys, we compare
the fusion probabilities at different injection points, where the
shell correction energy has very large negative and positive values,
such as $\alpha=0.46$ and $0.24$ in Fig.~2, respectively.

The former corresponds to the reaction $^{76}$Ge + $^{208}$Pb, which
is a cold fusion reaction, and the latter corresponds to $^{108}$Ru
+ $^{176}$Yb. Figure~3(a) shows sample trajectories that are
projected onto the $z-\alpha$ $(\delta=0)$ plane of the potential
energy surface for $^{284}114$. The contact point is marked by
$(+)$. The light yellow line denotes the trajectory with the
starting point $\alpha_{0}=0.46$ at the incident energy
corresponding to the excitation energy of the compound nucleus
$E^{*}$=20 MeV. This trajectory appears to move along the valley
till $z \sim 1.0$. Then it approaches the $z \sim 0$ region.

On the other hand, the trajectory with the starting point
$\alpha_{0}=0.24$ moves till $z \sim 1.0$ without changing the mass
asymmetry parameter $\alpha$, which is denoted by the white line. It
seems that the trajectory overcomes the mountain located at $z=1.4$
and $\alpha=0.24$. This trajectory does not move along the bottom of
the valley. We can say that this trajectory is not influenced by the
valley till $z \sim 1.0$. Then it turns in the $+z$ direction, which
corresponds to the fission region.


We project the trajectories of Fig.~3(a) onto the $z-\delta$ plane
at $\alpha=0$, as shown in Fig.~3(b).  The trajectory with
$\alpha_{0}=0.46$ can approach the compact shape region, but it is
not trapped by the pocket at the ground state. The trajectory with
$\alpha_{0}=0.24$ moves rapidly in the $+\delta$ direction and
enters the fission region. We can say that the former is a deep
quasi-fission process (DQF) and the latter is a quasi-fission
process (QF) \cite{ari04,ari05I}.

To understand why these two trajectories take different paths, we
precisely analyze the trajectory's behavior with the potential
energy surface near the contact point. The potential landscape near
the contact point is very important for deciding the destiny of the
trajectory \cite{ari05}. Figure~4(a) shows the trajectory with
$\alpha_{0}=0.46$, which is projected onto the $z-\delta$ plane
($\alpha=0.46$). The trajectory is the same as that in Fig.~3. The
dashed line denotes the ridge line. The contact point is far from
the ridge line, but it does not mean the trajectory moves into the
fission region immediately. On the potential energy surface near the
contact point, although we can see the steep slope in the $+\delta$
direction, there are no significant barriers in the $-z$ direction
from the contact point. Therefore, first the trajectory moves in the
$-z$ direction while maintaining $\delta \sim 0$. Owning to the
shell correction energy, a flat potential region exists around $z
\sim 0.8$ and $\delta \sim 0.25$. This flat region prevents the
trajectory from moving in the $+\delta$ direction rapidly. Finally,
the trajectory can enter the region near $z \sim 0$.

Figure~4(b) shows the trajectory with $\alpha_{0}=0.24$, which is
projected onto the $z-\delta$ plane ($\alpha=0.24$). The contact
point is located near the ridge line, but a substantial barrier
exists in the $-z$ direction from the contact point. To avoid the
barrier, the trajectory moves in the $+\delta$ direction rapidly.






\begin{figure}
\centerline{
\includegraphics[height=.40\textheight]{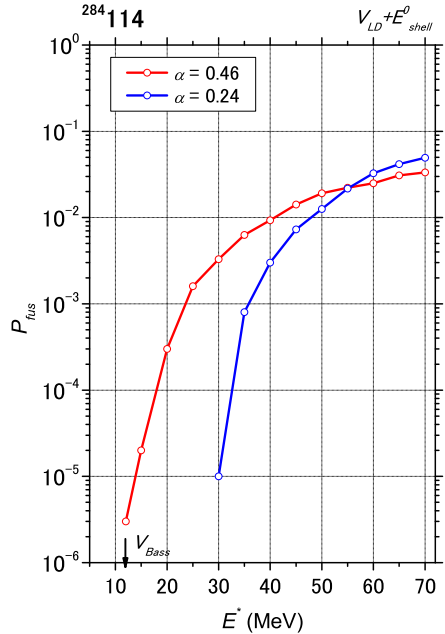}}
  \caption{Fusion probabilities for the initial values
$\alpha_{0}=0.46$ and $\alpha_{0}=0.24$, which are denoted by the
red and blue lines, respectively. }
\end{figure}

Figure~5 shows the fusion probabilities for $\alpha_{0}=0.46$ and
$\alpha_{0}=0.24$, which are denoted by the red and blue lines,
respectively. The potential energy at the contact point for
$\alpha=0.46$ is lower than that for $\alpha= 0.24$, as shown in
Fig.~2. Therefore, at low excitation energy, the former fusion
probability is larger than the latter one because in the former case
the available kinetic energy at the contact point is larger than
that in the latter case. The arrow indicates the Coulomb barrier
\cite{bass741}. At high excitation energy, the fusion probabilities
of these two cases are not markedly different because the kinetic
energy at the contact point is rather large in both cases.

%
%
%



\subsection{Evolution of cold fusion valleys on dynamical process}

Usually, cold fusion valleys are discussed using the potential
energy surface on the $z-\alpha$ plane with $\delta=0$, as in
Fig.~1(b). However, in the dynamical process, the trajectory moves
in the large $+\delta$ direction. Therefore, we should discuss the
cold fusion valleys with the dynamical evolution of the $\delta$
parameter. For the $\delta= 0$ case, we can see that the cold fusion
valleys are remarkable. With changing $\delta$ in the dynamical
process, the cold fusion valleys also change. While the trajectory
moves from $z \sim ~1.25$ to 0.5, $\delta$ changes to 0.2.
Figure~6(a) shows the potential energy surface on the $z-\alpha$
plane at $\delta=0.2$ with the trajectory. The position and depth of
the cold fusion valleys slightly change. After moving along the
valley, the trajectory enters the region $z < 0.5 $. In this region,
$\delta$ is more than 0.4. Figure~6(b) shows the potential energy at
$\delta=0.4$ with the trajectory. We can see that the cold fusion
valleys have already disappeared.


\begin{figure}
\centerline{
\includegraphics[height=0.30\textheight]{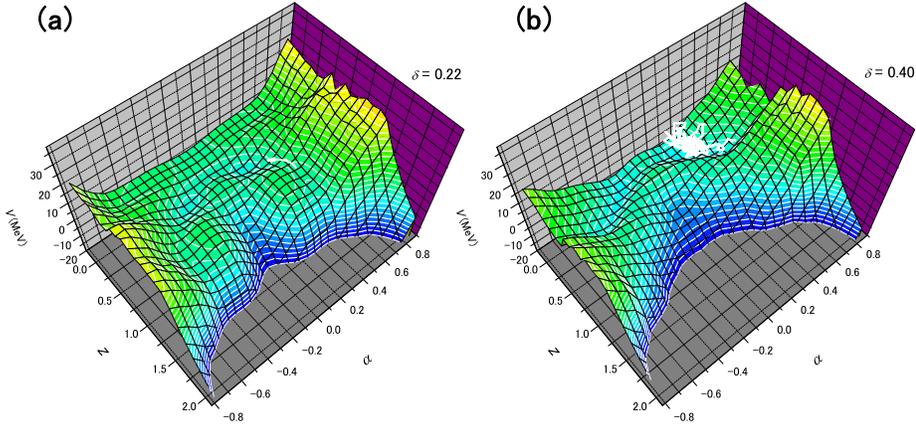}
}
  \caption{Potential energy surface $V_{LD}+E^{0}_{shell}$ in the $z-\alpha$ space
at (a) $\delta=0.2$ and (b) $\delta=0.4$ for $^{284}$114 with a
trajectory.}
\end{figure}

\subsection{Roles of shell effects on fusion process}

Owing to the shell correction energy, we discuss another important
effect to enhance the fusion probability. In the discussion on
fusion hindrance \cite{ari05I}, we indicated that the turning point
is very important. Figure~7(a) shows the potential energy surface of
$V_{LD}$ for $Z$=102 at the turning point on the $z-\delta$ plane,
that is to say, $\alpha$ corresponds to the value at the turning
point. The mean trajectory is denoted by the white line. The red
line denotes the trajectory taking into account the fluctuation. At
the turning point $(z \sim 0.3, \delta \sim 0.5)$, the trajectory
moves in the fission direction owing to the potential landscape
\cite{ari05I}.

However, when we take into account the shell correction energy shown
in Fig.~7(c), a temporary pocket appears at the turning point
(indicated by $A$), which we call the second pocket \cite{ari05}.
Figure~7(d) shows the potential energy at $\alpha=0$. The pocket
(indicated by $B$) corresponds to the ground state. The trajectory
is trapped in the second pocket $A$ at the turning point, and is
blocked from going to the fission area \cite{ari05}. While the
trajectory remains in pocket $A$, the mass asymmetry is relaxed, and
the large pocket $B$ appears at $\alpha \sim 0$. Then, the
trajectory moves to pocket $B$. Pocket $A$ helps the trajectory
enter the region corresponding to the compound nucleus.

Figure~7(b) shows the time evolution of nuclear shape. The
horizontal axis denotes $\delta$. After the contact point, the
trajectory moves in the large $\delta$ direction rapidly, then it is
trapped in the second pocket $A$. At this moment, the neck of
nuclear shape is rather large owing to the large $\delta$, so that
the mass asymmetry is relaxed easily while the trajectory remains in
pocket $A$. As $\alpha$ approaches zero, the trajectory moves into
pocket $B$. It corresponds to the compound nucleus.





\begin{figure}
\centerline{
\includegraphics[height=0.62\textheight]{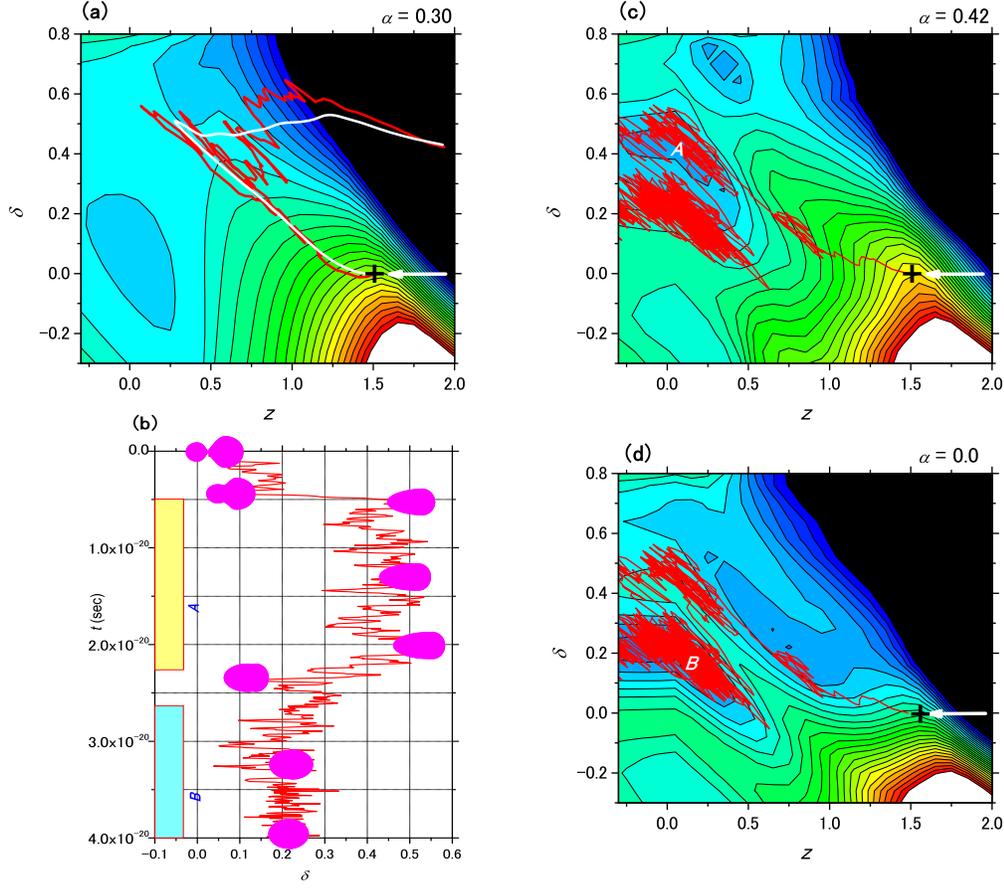}
}
  \caption{Sample trajectories projected onto the $z-\delta$ plane
at $\alpha$ which corresponds to the turning point. (a) $V_{LD}$
case and (c) $V_{LD}+E^{0}_{shell}$ case for $^{256}$No.
$V_{LD}+E^{0}_{shell}$ with $\alpha=0.0$ is shown in (d). The
evolution of nuclear shape is presented in (b).}
\end{figure}


\begin{figure}
\centerline{
\includegraphics[height=.38\textheight]{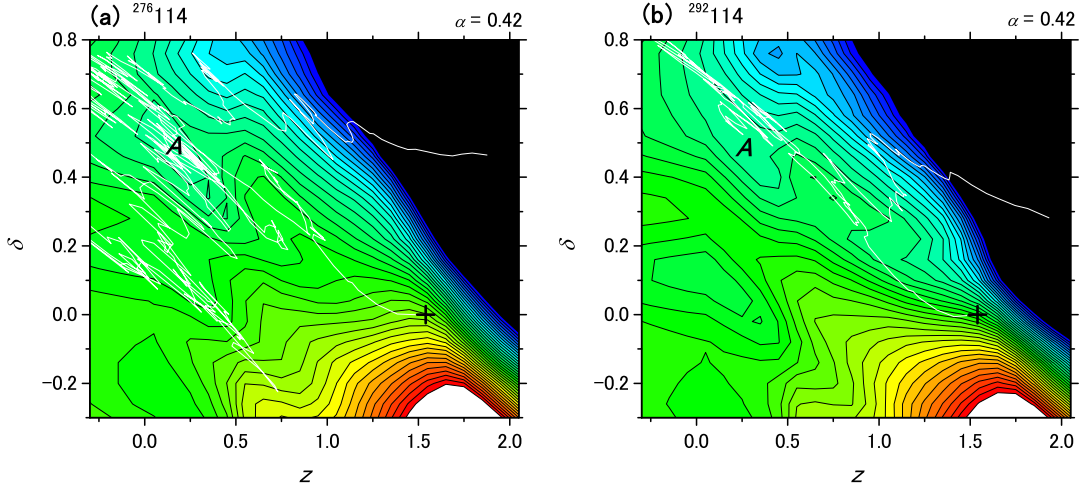}}
  \caption{Sample trajectory projected onto the $z-\delta$ plane
at $\alpha=0.42$ of $V_{LD}+E^{0}_{shell}$ for (a) $^{276}$114
(deformed nucleus) and (b) $^{292}$114 (spherical one). The
trajectory starts at $\alpha=0.67$ and $E^{*}=40$ MeV. Symbols are
given in the text.}
\end{figure}


\subsection{Difference between spherical and deformed compound nuclei}

In the previous section, we pointed out the role of the second
pocket in the fusion process. However, the second pocket does not
always enhance the fusion probability. As regards the position of
the second pocket, we discuss the difficulty of synthesizing
superheavy elements.

In general, the position at the ground state of the compound nucleus
is decided by the shell correction energy. As discussed above, in
the reaction $^{48}$Ca+ $^{208}$Pb, the ground state of the compound
nucleus $^{256}$No is located at $\delta \sim 0.2$, which is
deformed. Therefore, the second pocket $A$ is located at a $\delta$
with a more deformed shape $(\delta \sim 0.4)$, as shown in
Fig.~7(c), which simply corresponds to the turning point. In this
case, pocket $A$ assists the trajectory in entering the region
corresponding to the compound nucleus.

However, in the superheavy mass region $(Z > 110)$, the nuclear
shape at the ground state is spherical, which has a large shell
correction energy. For example, in the reaction $^{48}$Ca+
$^{244}$Pu $(\alpha=0.67)$, the ground state of the compound nucleus
$^{292}$114 is located at $\delta=0$. In this case, the second
pocket is located at $\delta \sim 0.2$, which does not correspond to
the turning point $(\delta \sim 0.4)$. Therefore, many trajectories
return in the $+z$ direction. This is one of the difficulties in
making spherical compound nuclei such as $^{292}$114.

To show the role of the second pocket, we compare the trajectories
forming the compound nuclei $^{276}114$ (deformed) and $^{292}114$
(spherical). Figures~8(a) and (b) show sample trajectories, which
are projected onto the $z-\delta$ plane at $\alpha=0.42$ of
$V_{LD}+E^{0}_{shell}$, for $^{276}114$ and $^{292}114$,
respectively. Both trajectories start at $\alpha=0.67$ with the
incident energy corresponding to the excitation energy of the
compound nucleus, $E^{*}$=40 MeV. In this case, $\alpha \sim 0.42$
corresponds to the turning point.

In Fig.~8(a), a small pocket exists at $z \sim 0.25$ and $\delta
\sim 0.5$, which is the second pocket (indicated by $A$). The
trajectory is trapped by pocket $A$ and is prevented from moving
into the reseparation region. Then, it moves to the main pocket,
which is located at around $z \sim 0, \delta \sim 0.2$ and $\alpha
\sim 0$. The shell correction energy at the ground state is about
5.0 MeV; therefore, the trajectory easily escapes from the main
pocket. In Fig.~8(b), which corresponds to the reaction
$^{48}$Ca+$^{244}$Pu, the second pocket does not exist at around $z
\sim 0.25$ and $\delta \sim 0.5$ because the compound nucleus
$^{292}$114 is spherical. We can see that it is difficult for the
trajectory to enter the spherical region. The trajectory returns to
the fission region.


Figure~9 shows the fusion probabilities for forming the compound
nuclei $^{292}114$ and $^{276}114$, which are denoted by the blue
and red lines, respectively. We use the initial value of
$\alpha=0.67$ for both cases. The fusion probability of $^{276}114$
is larger than that of $^{292}114$, owing to the second pocket,
which is located at the deformed shape.



\begin{figure}
\centerline{
\includegraphics[height=.40\textheight]{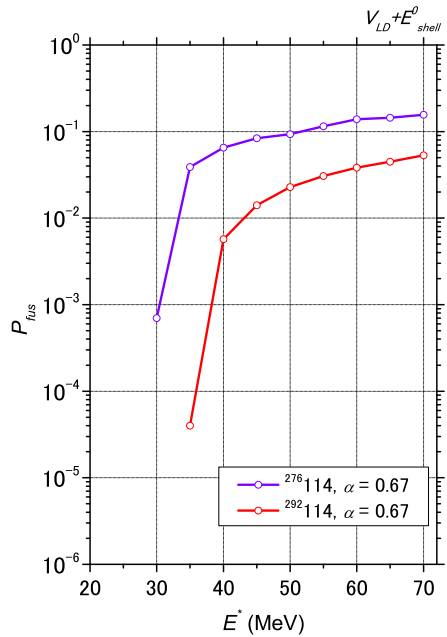}}
  \caption{Fusion probabilities for forming the compound nuclei
$^{292}114$ (spherical) and $^{276}114$ (deformed), which are
denoted by the blue and red lines, respectively.  We use the initial
value of $\alpha=0.67$ for both cases.}
\end{figure}


\section{Effect of temperature-dependent shell correction
energy}

In the previous section, we used the full shell correction energy
for the potential energy surface. However, the shell correction
energy actually depends on the nuclear temperature $T$. For the
dynamical calculation for the fusion process, the temperature
dependence of shell correction energy has not been considered
sufficiently in many publications.



In this section, we discuss the temperature dependence of the shell
correction energy and how the fusion process is affected.
Considerable effort has been made to investigate the temperature
dependence of the level density parameter \cite{igna75,igna79} and
applied to the calculation of the statistical model for the fission
process \cite{reis92}. The temperature dependence of the potential
energy surface has been investigated in references
\cite{dieb81,loje85}. Here, we discuss the effect of this dependence
on the fusion process on the basis of dynamical calculations.

\subsection{Temperature dependence of shell correction energy}

The temperature dependence of shell correction energy can be
extracted from the free energy calculated using single particle
energies, which is obtained from the two-center shell model code
\cite{suek74,iwam76}. The procedure is given precisely in reference
\cite{ari98}.
The temperature dependence of shell correction energy calculated by
this method is represented by $E_{shell}(q,T)$.


\begin{figure}
\centerline{
\includegraphics[height=.40\textheight]{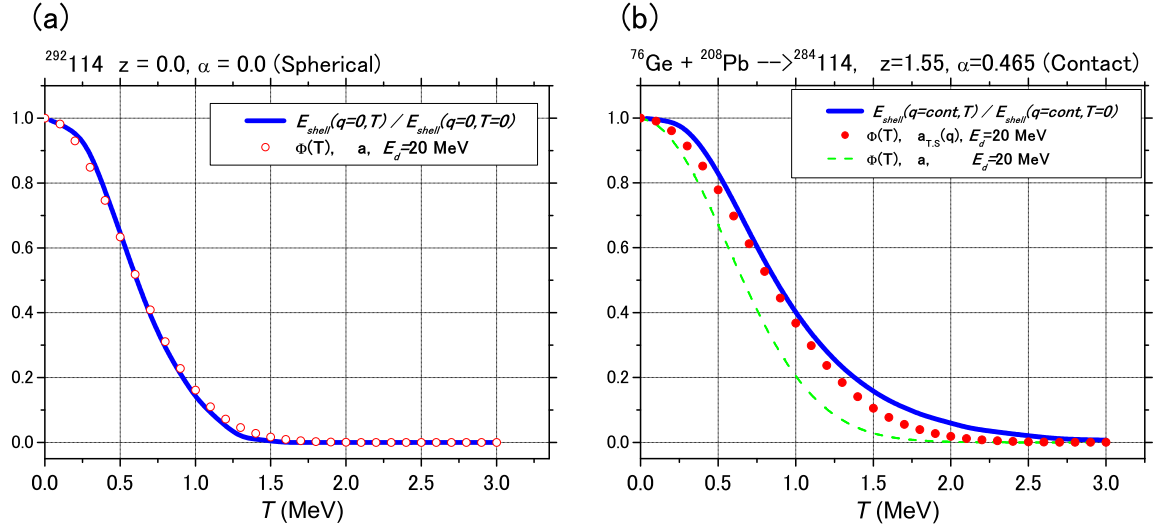}}
  \caption{Calculated temperature dependence of shell correction energy. (a)~For
  $^{292}114$ at $q=0$ (the spherical shape), compared with Ignatyuk's
  parameterization with $a=A/8$ and $E_{d}=20$ MeV. (b)~For $^{76}$Ge+$^{208}$Pb reaction
  at the contact point. The symbols and lines are explained in the text.}
\end{figure}

The ratio of shell correction energies,
$E_{shell}(q,T)/E_{shell}(q,T=0)$, in the spherical region for
$^{292}114$ is shown in Fig.~10(a) by the solid line. As discussed
in reference \cite{ari98}, our result is consistent with the
parameterization of the factor $\Phi(T)$,

\begin{eqnarray}
\Phi(T)=\exp \left( -\frac{aT^{2}}{E_{d}}\right),
\end{eqnarray}

following the work by Ignatyuk {\it et al.} \cite{igna75}, where $a$
denotes the level density parameter of the constant value $a=A/8$.
The shell damping energy $E_{d}$ is chosen as 20 MeV, which is given
by Ignatyuk {\it et al.} \cite{igna75}. This parameterization is
denoted by the open circles in Fig.~10(a), which shows a tendency
similar to that of the solid line.

In the fusion process, it is important to determine $E_{shell}(q,T)$
near the contact point of the colliding partner. The potential
landscape near the contact point mainly decides the destiny of the
trajectory in the dynamical process. Using the same procedure, we
calculate the temperature dependence of shell correction energy at
the contact point $E_{shell}(q=cont,T)$ for the reaction
$^{76}$Ge+$^{208}$Pb. In Fig.~10(b), the blue line denotes
$E_{shell}(q=cont,T)/E_{shell}(q=cont,T=0)$. This tendency is
different from that of the spherical shape case, which is denoted by
the dashed line. In Eq.~(7), when we take into account the nuclear
deformation dependence of the level density parameter $a_{T.S.}(q)$
\cite{toke81}, our result can be fitted using $E_{d}=20$ MeV,
denoted by the open circles. The result shows the similar tendency
of the blue line.



\begin{figure}
\centerline{
\includegraphics[height=.35\textheight]{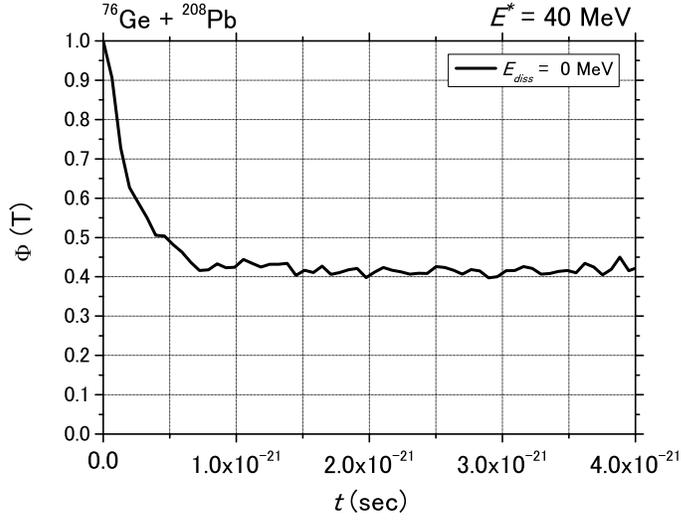}}
  \caption{Time evolution of $\Phi(T)$ for a trajectory in the reaction
$^{76}$Ge+$^{208}$Pb at $E^{*}=40$MeV with $E_{diss}=0$ MeV.}
\end{figure}


\begin{figure}
\centerline{
\includegraphics[height=.55\textheight]{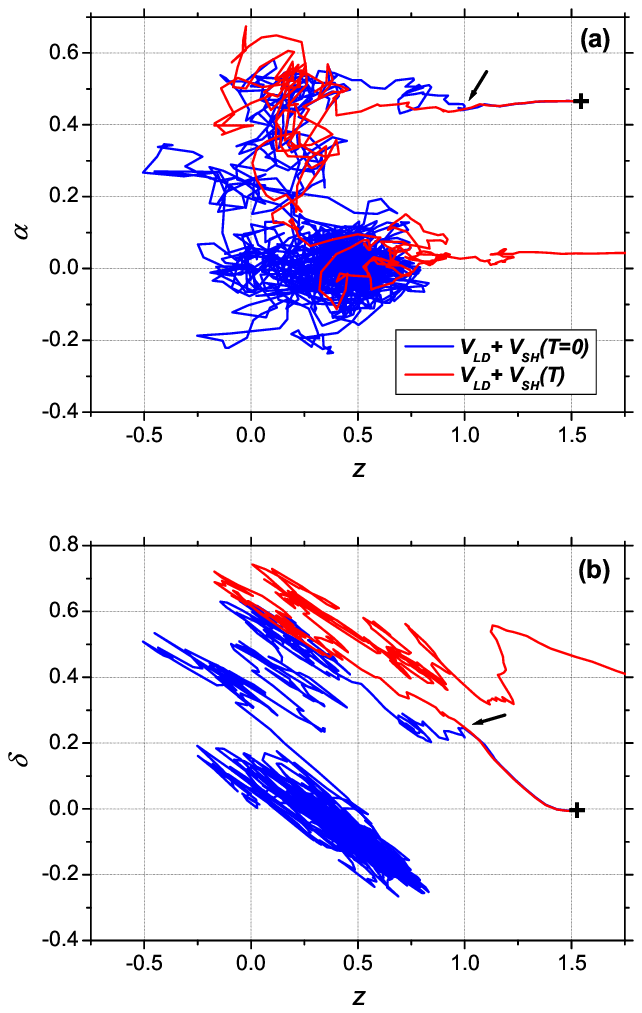}}
  \caption{Sample trajectory on $V_{LD}+V_{SH}(T=0)$ and $V_{LD}+V_{SH}(T)$
  in the reaction $^{76}$Ge+$^{208}$Pb at $E^{*}=40$ MeV, which are denoted by blue
and red lines, respectively. The trajectories are projected onto (a)
the $z-\alpha$ plane and (b) the $z-\delta$ plane, respectively.
Symbols are given in the text.}
\end{figure}

\subsection{Trajectory behavior on temperature-dependent potential energy surface}

In the fission process, the time evolution of the potential energy
surface affects the dynamical process considerably. Owing to the
neutron emission from the compound nucleus, the temperature of the
nucleus decreases, and the shell correction energy recovers. The
time scale of the variation of the potential energy surface affects
the dynamics of the fission process \cite{ari98}.

Here, we focus on the effect of the temperature-dependent shell
correction energy on the fusion process. In the dynamical
calculation, we analyze the trajectory on the temperature-dependent
potential energy surface, which is concerned with the intrinsic
energy $E_{int}$ of the system.

Within our model, the time evolution of $E_{int}$ is controlled by
the friction tensor $\gamma$. Also, $E_{int}$ is connected with
$E_{diss}$ at the contact point \cite{ari052}. Here, we clarify the
relationship among these variables and the mechanism of the fusion
process.

We start the dynamical calculation from the contact point using the
Langevin equation. $E_{diss}$ is treated as a parameter of the
initial condition. Discussions on this issue have been given in
references \cite{ari03a,ari052}. In fact, $E_{diss}$ is related to
the dynamics in the approaching process. Considering $E_{diss}$ as a
parameter, we investigate the effect of $E_{diss}$ on the trajectory
in the process after the both nuclei touch. Then, in the next
section, we discuss the effect of the friction tensor $\gamma$.

Before we discuss the effect of the different $E_{diss}$'s on the
dynamical process, we analyze the trajectory under the influence of
the temperature-dependent shell correction energy, using the same
initial condition as $E_{diss}=0$ MeV in the previous calculation.
We employ the potential energy surface $V_{LD}+V_{SH}(T=0)$ and
$V_{LD}+V_{SH}(T)$ in the reaction $^{76}$Ge+$^{208}$Pb.

When we use $V_{LD}+V_{SH}(T)$ as the potential energy surface, as
the intrinsic energy that is transformed from kinetic energy
increases, some part of the shell correction energy disappears.
Figure~11 shows the time evolution of $\Phi(T)$ for a trajectory at
$E^{*}=40$ MeV. At the contact point, where $t=0$ in the
calculation, owing to the assumption $E_{diss}=0$ MeV, which
corresponds to $E_{int}=0$ MeV, the full shell correction energy is
indicated. Then, till $t \sim 1.0 \times 10^{-21}$ sec, more than
50\% of the shell correction energy disappears owing to the strong
friction tensor. During this extremely short reaction time, the
kinetic energy dissipates into the intrinsic energy, which has been
discussed in reference \cite{ari05}.


\begin{figure}
\centerline{
\includegraphics[height=.40\textheight]{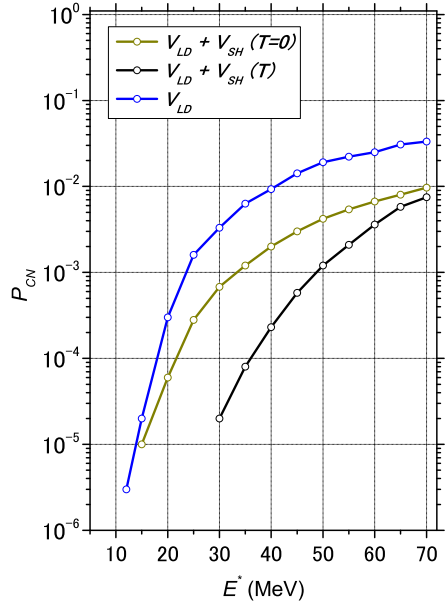}}
  \caption{Fusion probabilities $P_{CN}$ in the reaction
$^{76}$Ge+$^{208}$Pb for each potential energy with $E_{diss}=0$
MeV. The blue and red lines denote $P_{CN}$ with
$V_{LD}+V_{SH}(T=0)$ and $V_{LD}+V_{SH}(T)$, respectively. $P_{CN}$
with $V_{LD}$ is represented by the black line.}
\end{figure}

As consequence, such variation of the potential energy surface gives
the influence on the trajectory's behavior. Figure~12 shows sample
trajectories on $V_{LD}+V_{SH}(T=0)$ and $V_{LD}+V_{SH}(T)$, which
are denoted by the blue and red lines. The trajectories are
projected onto the $z-\alpha$ plane and $z-\delta$ plane, which are
shown in (a) and (b), respectively. The (+) denotes the point of
contact in the system. Both trajectories in Fig.~12 are calculated
using the same random numbers. After $t=7.9 \times 10^{-22}$ sec,
which is indicated by the black arrow, trajectories separate along
different paths owing to the different potential energy surfaces.
The trajectory with $V_{LD}+V_{SH}(T=0)$ takes the fusion process.
On the other hand, the trajectory with $V_{LD}+V_{SH}(T)$ takes the
deep quasi-fission process. This is brought about by the
disappearance of the second pocket at around $z \sim 0.2, \delta
\sim 0.4$ and $\alpha \sim 0.3$.

Figure~13 shows the fusion probabilities $P_{CN}$ in the reaction
$^{76}$Ge+$^{208}$Pb for each potential energy with $E_{diss}=0$
MeV. The blue and red lines denote $P_{CN}$ with
$V_{LD}+V_{SH}(T=0)$ and $V_{LD}+V_{SH}(T)$, respectively. $P_{CN}$
with $V_{LD}$ is represented by the black line. With increasing
incident energy, $P_{CN}$ with $V_{LD}+V_{SH}(T)$ approaches that
with $V_{LD}$ owing to the disappearance of shell correction energy
at high temperature.

The mass distributions of fission fragments in the reaction
$^{76}$Ge+$^{208}$Pb at $E^{*}=40$ MeV are shown in Fig.~14. The
blue and red lines denote the cases of $V_{LD}+V_{SH}(T=0)$ and
$V_{LD}+V_{SH}(T)$ with $E_{diss}=0$ MeV, respectively. In the
latter case, some part of the shell correction energy disappears and
the number of mass symmetric fission events increases in comparison
with that in the $V_{LD}+V_{SH}(T=0)$ case.



\begin{figure}
\centerline{
\includegraphics[height=.36\textheight]{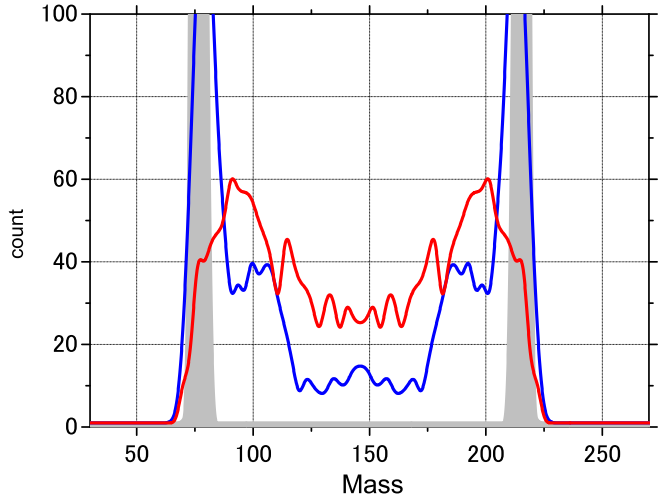}}
 \caption{Mass distribution of fission
fragments in the reaction $^{76}$Ge+$^{208}$Pb at $E^{*}=40$ MeV.
The blue and red lines denote the cases of $V_{LD}+V_{SH}(T=0)$ and
$V_{LD}+V_{SH}(T)$ with $E_{diss}=0$\%, respectively.  The gray
shadow denotes the case of $V_{LD}+V_{SH}(T)$ with $E_{diss}=100$\%.
}
\end{figure}

\subsection{$E_{diss}$ dependence on dynamical process}

Next, we focus on the dependence on $E_{diss}$. As discussed in
references \cite{ari052,ari03a}, the fusion probability and mass
distribution of fission fragments are affected by $E_{diss}$. When
we use the condition  $E_{diss}=0$ MeV (or $E_{diss}=0 \%$), we
assume that the velocity of the trajectory is in the $-z$ direction
at the contact point, so that there exists trajectories that enter
the fusion region. However, when we take the condition in which all
the kinetic energy has already dissipated at the contact point,
which is represented by $E_{diss}=100\%$, all trajectories move down
along the steep slope in the fission direction (the $+z$ direction)
immediately. As a result, the fusion probability is equal to zero.
We discuss this point in connection with the potential energy
surface near the contact point in section 4.5.


\begin{figure}
\centerline{
\includegraphics[height=.40\textheight]{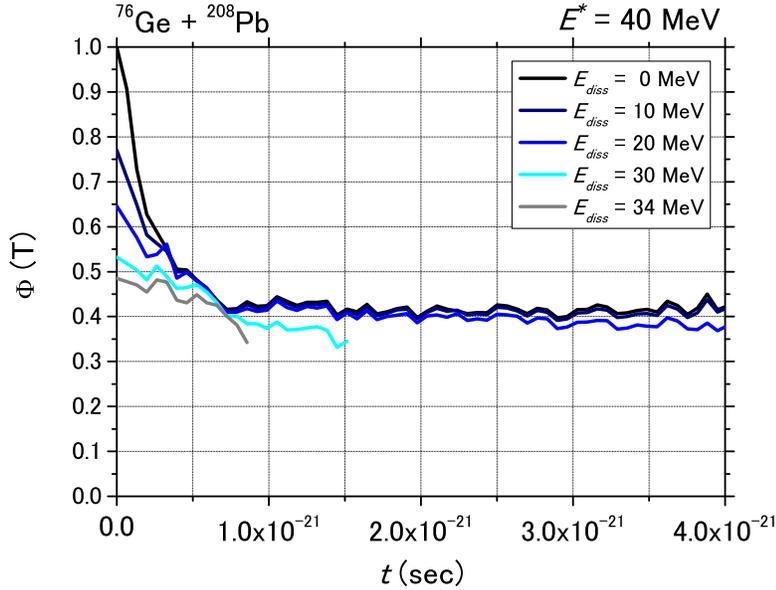}}
  \caption{Time evolution of $\Phi(T)$ for a trajectory in the
reaction $^{76}$Ge+$^{208}$Pb at $E^{*}=40$ MeV for various
$E_{diss}$ cases.}
\end{figure}

First, we investigate the evolution of $\Phi(T)$ related to
$E_{diss}$. When the kinetic energy dissipates during the
approaching process, some part of the shell correction energy has
already disappeared at the contact point. Figure~15 shows the time
evolution of $\Phi(T)$ for a trajectory at $E^{*}=40$ MeV with
various $E_{diss}$. At $t=0$ sec, $\Phi(T)$ strongly depends on
$E_{diss}$. In this case, the potential energy surface at the
contact point for each $E_{diss}$ is shown in Fig.~16.
Owing to the strong friction,  $60 \%$ of the shell correction
energy is rapidly lost by $7.5 \times 10^{-22}$ sec. After this
time, the trajectory of each $E_{diss}$ moves on almost the same
potential energy surface.

In Fig.~15, the gray line denotes $\Phi(T)$ for $E_{diss}=100\%$,
which corresponds to $E_{diss}=34$ MeV in this case. In fact, in the
surface friction model, it shows that all the kinetic energy
dissipates during the approaching process \cite{gros78,froe83}.
Under this condition, the gray line shows that 50\% of the shell
correction energy has already disappeared at the contact point. The
trajectory moves down the steep slope rapidly and arrives at the
fission region at $8.0 \times 10^{-22}$ sec. The mass distribution
of the fission fragments in this reaction is denoted by the gray
shadow in Fig.~14. As discussed in reference \cite{ari03a}, it shows
the two sharp peaks corresponding to target-like and projectile-like
fragments.


The fusion probabilities for the various $E_{diss}$ cases are shown
in Fig.~17. We can roughly say that the fusion probability of each
$E_{diss}$ shifts by as much as $E_{diss}$ MeV from the black line
in the $E_{diss}=0$ MeV case. For example, in Fig.~17, we connect
the points that have the same value of $E^{*}-E_{diss}$ at the
contact point using the red dashed line. On this dashed line, the
fusion probability with the large $E_{diss}$ increases owing to the
thermal fluctuation of the collective motion. However, we can see
the systematic tendency and the relationship between the fusion
probability and $E_{diss}$. Even though we treat $E_{diss}$ as a
parameter in our model, the mechanism of the dynamical process after
the contact point is clarified.


\begin{figure}
\centerline{
\includegraphics[height=.40\textheight]{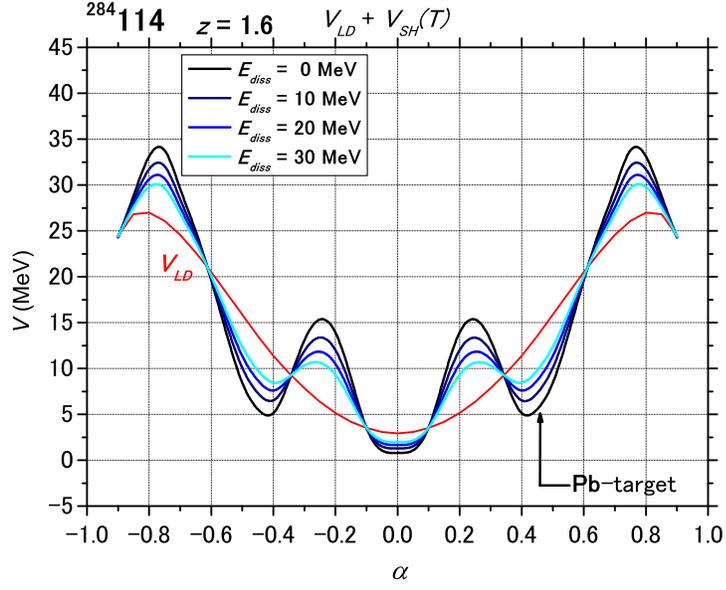}}
  \caption{Potential energy surface $V_{LD}+V_{SH}(T)$ at the contact
  point in the reaction $^{76}$Ge+$^{208}$Pb at $E^{*}=40$ MeV
  for various $E_{diss}$ cases. }
\end{figure}


\begin{figure}
\centerline{
\includegraphics[height=.40\textheight]{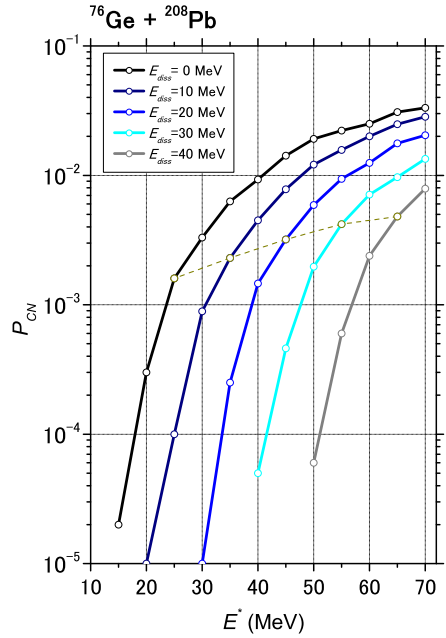}}
  \caption{Fusion probabilities $P_{CN}$ in the reaction
$^{76}$Ge+$^{208}$Pb with $V_{LD}+V_{SH}(T)$ for various $E_{diss}$
cases.}
\end{figure}

\subsection{Friction tensor dependence on dynamical process}

In the dynamical process, the dissipation speed is controlled by the
friction tensor $\gamma$. That is to say, the time evolution of
$E_{int}$ depends on $\gamma$. The ambiguity of the friction tensor
derived from the macroscopic model has been discussed very often
\cite{wada92,hils93,karp01}. Moreover, as for the effect of nuclear
structure, we should consider the microscopic transport
coefficients, that are calculated by the linear response theory
\cite{ivan97,hofm97,yama97}. The friction tensor that calculated by
the microscopic model depends on the temperature and is smaller than
that calculated by the macroscopic model at low temperature. Here,
using the modified strength of the one-body friction tensor, we
investigate the effect of the friction tensor on the dynamical
process, in terms of the variational speed of the potential energy
surface.


\begin{figure}
\centerline{
\includegraphics[height=.40\textheight]{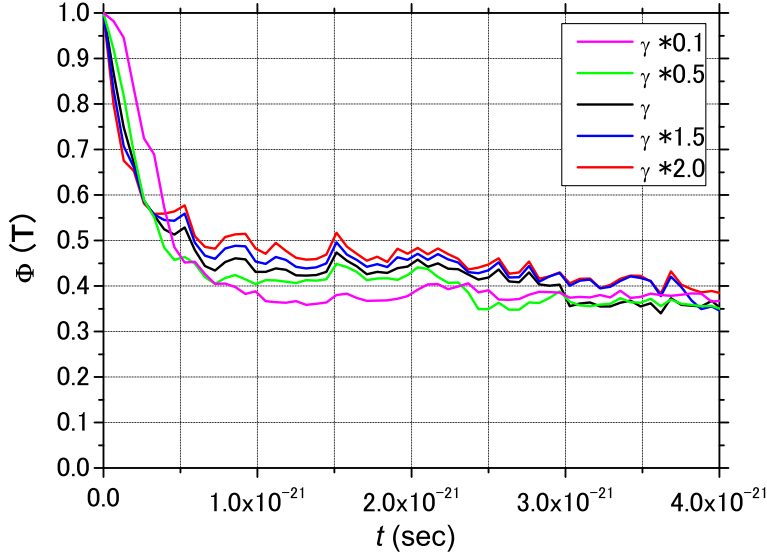}}
  \caption{Time evolution of $\Phi(T)$
for the different friction tensors $\gamma$ in the reaction
$^{76}$Ge+$^{208}$Pb at $E^{*}=40$ MeV with $E_{diss}=0$ MeV.}
\end{figure}

Figure~18 shows the time evolution of $\Phi(T)$ for the different
friction tensors in the reaction $^{76}$Ge+$^{208}$Pb at $E^{*}=40$
MeV with $E_{diss}=0$ MeV. The factors multiplied with the friction
tensor are 0.1, 0.5, 1.0, 1.5 and 2.0, and the corresponding results
are shown. Except in the case of the multiplied factor 0.1, more
than  40\% of the shell correction energy disappears till $3.0
\times 10^{-22}$ sec for all cases. This time is rather short in
comparison with the reaction time in the whole dynamical process.
After $t=5.0 \times 10^{-22}$ sec, in the smaller friction case,
$\Phi(T)$ decreases faster than in the larger friction case. In the
smaller friction case, after the trajectory overcomes the fusion
barrier rapidly, the trajectory enters the region where the
potential energy is lower than that near the contact point. It is
possible that the variation in the potential energy is transformed
into the intrinsic energy by Eq.~(6). In this way, after the
trajectory overcomes the fusion barrier at high speed, the intrinsic
energy increases and the shell correction energy disappears rapidly.
The intrinsic energy increases, not only owing to the dissipation
from kinetic energy, but also owing to the transformation of the
variation from the potential energy.




\begin{figure}
\centerline{
\includegraphics[height=.40\textheight]{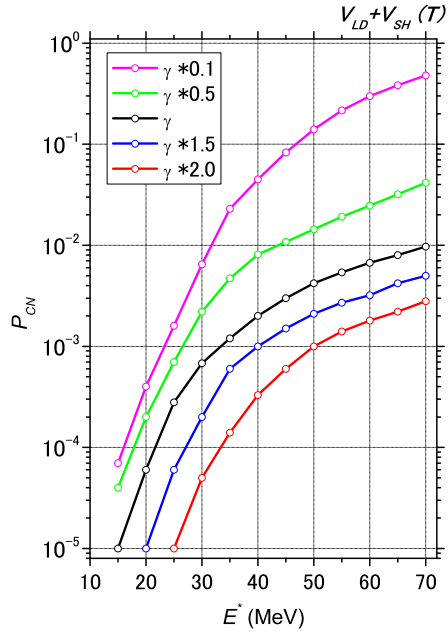}}
  \caption{Fusion probabilities $P_{CN}$ for the different friction
tensor $\gamma$ in the reaction $^{76}$Ge+$^{208}$Pb with
$E_{diss}=0$ MeV. $V_{LD}+V_{SH}(T)$ is employed. }
\end{figure}



Figure~19 shows the fusion probabilities $P_{CN}$ for each friction
tensor for the reaction $^{76}$Ge+$^{208}$Pb with $E_{diss}=0$ MeV.
$P_{CN}$ increases with decreasing magnitude of the friction tensor.
The friction tensor affects the dynamical process considerably
\cite{ari052}.

\begin{figure}
\centerline{
\includegraphics[height=.40\textheight]{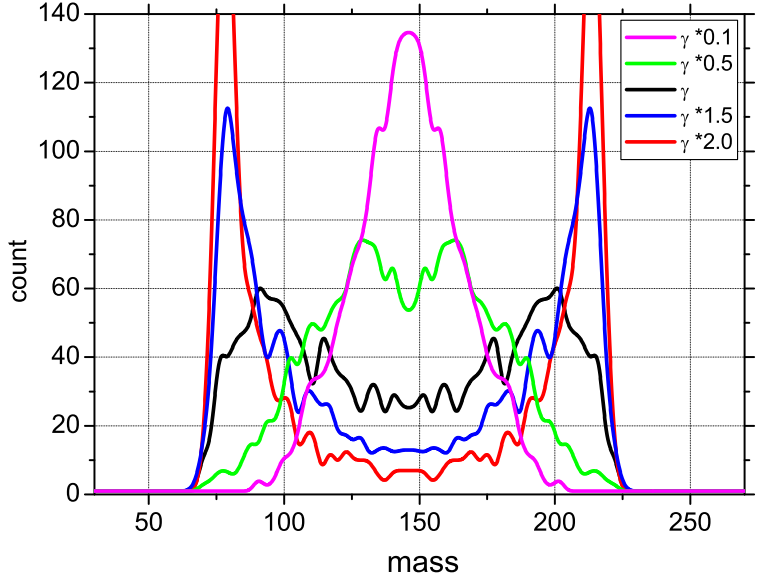}}
  \caption{Mass distribution of fission fragments for each
friction tensor $\gamma$ in the reaction $^{76}$Ge+$^{208}$Pb at
$E^{*}=40$ MeV with $E_{diss}=0$ MeV}
\end{figure}

We investigate the mass distribution of fission fragments for each
friction tensor for the reaction $^{76}$Ge+$^{208}$Pb at $E^{*}=40$
MeV with $E_{diss}=0$ MeV, which is shown in Fig.~20. In the small
friction case, mass symmetric fission events are dominant, because
the trajectory can reach the small $z$ region. On the other hand,
mass asymmetric fission events are dominant in the large friction
case. The smallest value of $z$ that the trajectory can reach is
related to the mass distribution of fission fragments \cite{ari05}.

\subsection{Effect of modified potential energy surface}

As we mentioned in section 4.3, when all the kinetic energy
dissipates during the approaching process $(E_{diss}=100\%)$, all
trajectories move down the steep slope in the fission direction and
the mass distribution of fission fragments has two sharp peaks, as
shown in Fig.~14. In connection with this problem, we should
precisely investigate the landscape of the potential energy surface
near the contact point. The modification of the potential energy
near the contact point is performed continuously \cite{zag05}.


\begin{figure}
\centerline{
\includegraphics[height=.40\textheight]{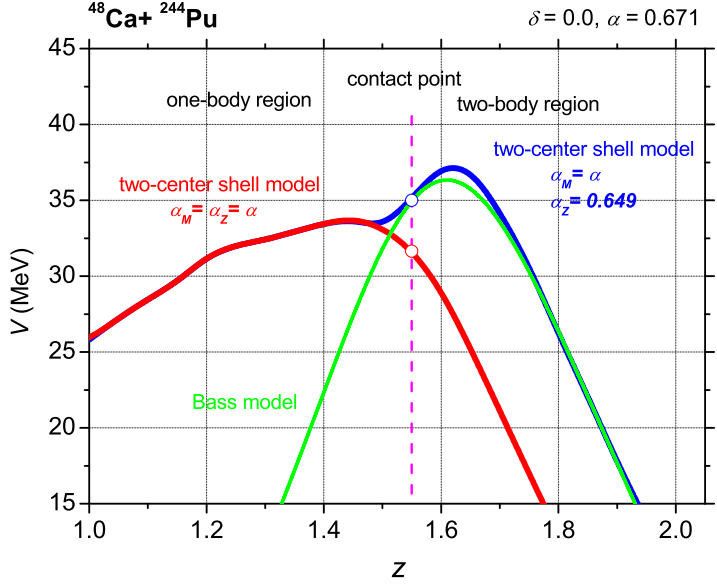}}
  \caption{Potential energy in the reaction
  $^{48}$Ca+$^{244}$Pu calculated using the two-center shell model. The
  red line denotes the potential energy using $\alpha_{M}=\alpha_{Z}$
  for all region. The blue line denotes
  the modified potential $V^{mod}$ using $\alpha_{M}$ and
  $\alpha_{Z}$ of the projectile and target in the two-body
  region. The potential energy calculated using the Bass model is
denoted by the green line. Symbols denote in the text.}
\end{figure}


\begin{figure}
\centerline{
\includegraphics[height=.40\textheight]{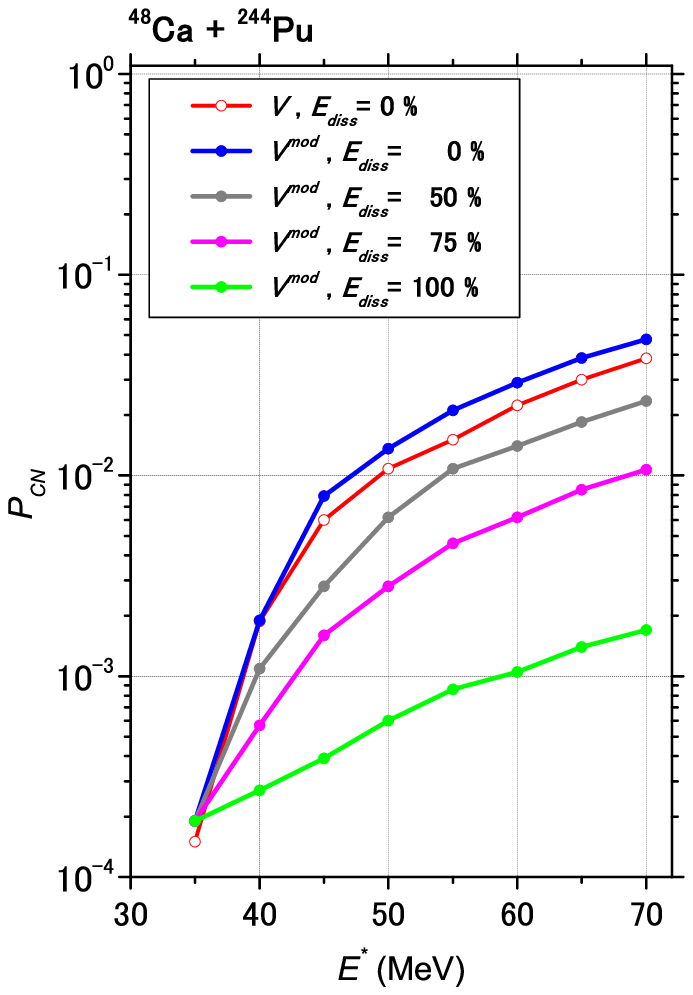}}
  \caption{Fusion probabilities $P_{CN}$ in the reaction
  $^{48}$Ca+$^{244}$Pu for each potential. In the case of $E_{diss}=0$ MeV,
$P_{CN}$ using the previous potential energy $V$ is denoted by the
red line. $P_{CN}$ using the modified potential energy $V^{mod}$ for
the different $E_{diss}$ are denoted by each line.}
\end{figure}

As a problem in our two-center shell model, since we employ the mass
asymmetry parameter $\alpha$, the charge distribution of the nuclei
follows $\alpha$. In the fusion process, this is not the proper
assumption before the contact point, which we call the two-body
region. At least, in the two-body region, we should calculate the
potential energy using realistic distributions for the proton charge
$\alpha_{Z}$ and  mass $\alpha_{M}$ of the projectile and target. By
performing the analysis in this manner in the two-body region, we
obtain the modified potential energy surface by employing the proper
smoothing between the two-body region and the one-body region at the
contact point, which is denoted by the blue line in Fig.~21. It
shows the $^{48}$Ca+$^{244}$Pu reaction case, where both colliding
nuclei are assumed to have a spherical shape. The red line denotes
the potential energy using the same distributions of $\alpha$ for
the charge and mass of both nuclei, even in the two-body region,
that were used in the previous calculation. The open circles denote
the contact point of the colliding partner. The Coulomb barrier of
the modified potential energy agrees with that calculated using the
Bass model, which is denoted by the green line \cite{bass741}.

Using the modified potential energy surface, which is represented by
$V^{mod}$, we calculate the fusion probability and mass distribution
of fission fragments using the same procedure as in the previous
section. We start the Langevin calculation at the contact point. We
also take into account the temperature dependence of shell
correction energy.

In the case of $E_{diss}=0$ MeV, the fusion probability with the
modified potential $V^{mod}$ does not change considerably compared
with the previous results. Since the velocity of the trajectory is
in the $-z$ direction, it is easy to enter the small $z$ region, and
it is not affected by the potential energy surface in the two-body
region. In the case of $E_{diss}=0$ MeV, Fig.~22 shows $P_{CN}$
using the previous potential and the modified potential, which are
denoted by the red and blue lines, respectively.

On the other hand, when all the kinetic energy dissipates during the
approaching process, even though the fusion probability with the
previous potential is zero for any incident energy, it markedly
increases in the case of the modified potential $V^{mod}$, which is
denoted by the green line in Fig.~22. In the $V^{mod}$ case, the
Coulomb barrier located at $z=1.63$ prevents the trajectory from
moving into the fission region immediately, and the trajectory at
the contact point moves down along the potential slope in the $-z$
direction automatically. In this case, the trajectory has the
momentum to go to the fusion region. In Fig.~22, we also show
$P_{CN}$ in the $E_{diss}=50 \%$ and 75 \% cases with $V^{mod}$.
$P_{CN}$ decreases with increasing $E_{diss}$.



\begin{figure}
\centerline{
\includegraphics[height=.40\textheight]{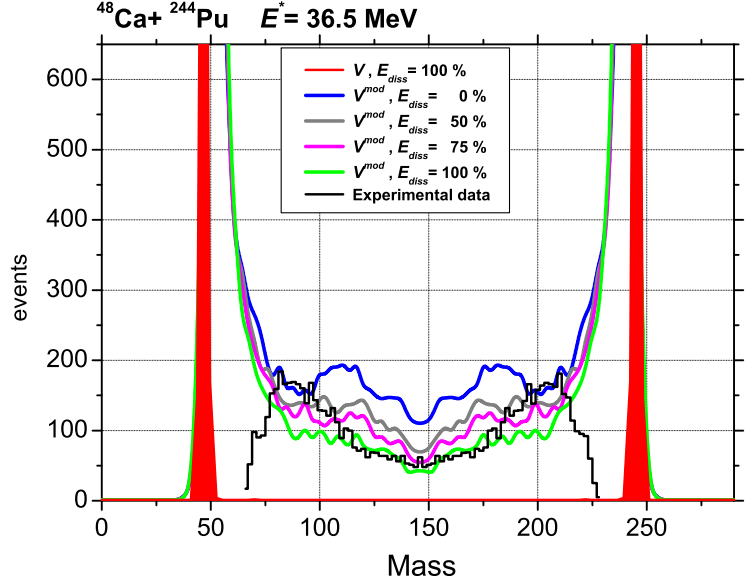}}
  \caption{Mass distributions of fission fragments with
the modified potential energy $V^{mod}$ in the reaction
$^{48}$Ca+$^{244}$Pu at $E^{*}=36.5$ MeV for the different
$E_{diss}$. The red shadow denotes the mass distributions using the
previous potential energy with $E_{diss}=0$ MeV. The experimental
data is denoted by black line \cite{itki01}.}
\end{figure}

Figure 23 shows the mass distribution of fission fragments with the
modified potential energy surface, in the reaction
$^{48}$Ca+$^{244}$Pu at $E^{*}=36.5$ MeV.
When all the kinetic energy dissipates during the approaching
process, the results with the previous and  modified potential
energy surfaces are denoted by the red shadow and green lines,
respectively. Whereas the result with the previous potential shows
two sharp peaks, that with the modified potential is distributed
around the wide mass region. The tendency of the latter shows a
rather good agreement with the experimental data, which correspond
to fusion-fission and quasi-fission events, as denoted by the black
line \cite{itki01}. Here, the experimental data are normalized to
agree with the results in the  $V^{mod}$ case with $E_{diss}=100 \%$
at the mass symmetric point. Also, we plot the results of different
$E_{diss}$ cases in Fig.~23.


Even though this result is a preliminary one, it shows the
importance of the potential energy surface near the contact point
and $E_{diss}$. We should investigate more precisely the effect of
the approaching process with the modified potential energy.



\section{Summary}

The dynamical process in the superheavy mass region was studied
systematically by trajectory calculation in three-dimensional
coordinate space. To understand the fusion mechanism clearly, we
treated the dynamical process as a fusion hindrance and as a fusion
enhancement, separately.

On the basis of our previous results \cite{ari05I}, we discussed the
fusion enhancement. We introduced shell correction energy on the
potential energy surface, and investigated how trajectory is
affected by shell correction energy.

The fusion probability increases along the cold fusion valleys at
low incident energy. This is caused by the potential energy surface
near the contact point on the $z-\delta$ plane. With shell
correction energy, the second pocket appears at the location with a
deformed shape. When the position of the second pocket corresponds
to the turning point, it enhances the fusion probability.

The temperature dependence of shell correction energy was also
discussed. In the dynamical process, the evolution of $\Phi(T)$ is
related to $E_{diss}$ and the friction tensor $\gamma$. We analyzed
the trajectory on the temperature-dependent potential energy
surface, and clarified the effects of $E_{diss}$ and $\gamma$. The
fusion probability and mass distribution of fission fragments were
calculated.

This study is the first systematic attempt to investigate the role
of shell correction energy in the dynamical process. Shell
correction energy plays a very important role in the fusion process
and enhances the fusion probability, when the colliding partner has
a strong shell structure. Although we treated $E_{diss}$ as a
parameter, after the both nuclei touch, the mechanism of the
dynamical process and the relationship among the parameters are
clarified.

As a topic of further study, the potential energy surface should be
improved near the contact point and we should investigate the
dynamical process more accurately \cite{zag05}. We should take into
account the dynamics in the approaching process and clarify how the
kinetic energy dissipates in the approaching process. Also, we
should directly apply the transport coefficients calculated using
the microscopic model, for example, linear response theory
\cite{ivan97,hofm97,yama97}.


The author is grateful to Professor M.~Ohta, Professor
Yu.~Ts.~Oganessian, Professor M.G.~Itkis, Professor V.I.~Zagrebaev,
Professor F.~Hanappe, Dr. A.K.~Nasirov and Dr. F.A.~Ivanyuk for
their helpful suggestions and valuable discussion throughout the
present work. The special thanks are devoted to Professor T.~Wada
who developed the original calculation code for the
three-dimensional Langevin equation. The author thanks Dr. S.~Yamaji
and his collaborators, who developed the calculation code for
potential energy with two-center parameterization.

This work has been in part supported by INTAS projects 03-01-6417.





\bibliographystyle{aipproc}   

\bibliography{sample}

\IfFileExists{\jobname.bbl}{}
 {\typeout{}
  \typeout{******************************************}
  \typeout{** Please run "bibtex \jobname" to optain}
  \typeout{** the bibliography and then re-run LaTeX}
  \typeout{** twice to fix the references!}
  \typeout{******************************************}
  \typeout{}
 }

\end{document}